\documentclass[letterpaper,times]{IONconf}
\usepackage[pdftex]{graphicx}
\DeclareGraphicsExtensions{.pdf,.jpeg,.png}

\usepackage{amsmath}
\usepackage{amssymb}

\usepackage[ruled]{algorithm2e}
\usepackage{algpseudocode}

\usepackage{array}
\usepackage{textcomp}
\usepackage{makecell}
\usepackage{hhline}

\usepackage{multicol}
\usepackage{multirow}
\usepackage{booktabs}
\usepackage{threeparttable}

\usepackage{float}

\usepackage{verbatim}

\usepackage{xcolor}

\usepackage{pifont}

\newcolumntype{P}[1]{>{\centering\arraybackslash}p{#1}}

\newcommand{\thickhline}{
    \noalign {\ifnum 0=`}\fi \hrule height 1pt
    \futurelet \reserved@a \@xhline
}

\makeatletter
\def\env@dmatrix{\hskip -\arraycolsep
  \let\@ifnextchar\new@ifnextchar
  \extrarowheight=2ex
  \array{*\c@MaxMatrixCols{>{\displaystyle}c}}}

\makeatother

\DeclareCaptionLabelSeparator{custom}{ }
\DeclareCaptionFormat{custom}
{%
    \textbf{#1#2}\textit{\small #3}
}
\usepackage{url}

\usepackage{natbib}

\usepackage[hidelinks]{hyperref}

\title{Set-Based Position Ambiguity Reduction Method for Zonotope Shadow Matching in Urban Areas Using Estimated Multipath Errors}

\author{
    Sanghyun~Kim and Jiwon~Seo, \textit{Yonsei~University}
    }

\begin{document}

\maketitle

\section*{Biography} 


\biography{Sanghyun Kim}{is a Ph.D. candidate in the School of Integrated Technology, Yonsei University, Incheon, South Korea. He received the B.S. degree in Integrated Technology from Yonsei University. His research interests include seamless positioning in urban environments and intelligent transportation systems.}

\biography{Jiwon Seo}{is an Underwood Distinguished Professor at Yonsei University, where he is a professor in the School of Integrated Technology, Incheon, South Korea. 
He also serves as an Adjunct Professor in the Department of Convergence IT Engineering at Pohang University of Science and Technology (POSTECH), Pohang, South Korea.
His research interests include GNSS anti-jamming technologies, complementary PNT systems, and intelligent unmanned systems.
Dr. Seo is a member of the International Advisory Council of the Resilient Navigation and Timing Foundation, Alexandria, VA, USA, and the Advisory Committee on Defense of the Presidential Advisory Council on Science and Technology, South Korea.}

\section*{Abstract}

In urban areas, the quality of global navigation satellite system (GNSS) signals deteriorates, leading to reduced positioning accuracy. 
To address this issue, 3D-mapping-aided (3DMA) techniques, such as shadow matching and zonotope shadow matching (ZSM), have been proposed. 
However, these methods can introduce a problem known as multi-modal position ambiguity, making it challenging to select the exact mode in which the receiver is located. 
Accurately selecting the correct mode is essential for improving positioning accuracy. 
A previous study proposed a method that uses satellite-pseudorange consistency (SPC), calculated from pseudorange measurements, to select the mode containing the receiver. 
This method achieved a mode selection accuracy of approximately 78\%. 
To further enhance accuracy, the study utilized pseudorange measurements collected at multiple timesteps from a fixed location and a trained line-of-sight (LOS) classifier. 
However, in practice, collecting data at multiple timesteps from the same location in dynamic environments is challenging. 
Moreover, the performance of the trained LOS classifier heavily depends on the surrounding environment, leading to low reliability. 
In this study, we propose a method that estimates and corrects multipath errors based on the mode distribution obtained from the output of ZSM and extract an enhanced SPC using the corrected pseudorange measurements. 
This enables high mode selection accuracy using only single-timestep pseudorange measurements, without requiring a trained LOS classifier. 
Experimental results using global positioning system (GPS) data collected in an urban environment demonstrate that the proposed method achieves a mode selection accuracy of 91\%, compared to 86\% for the existing method.
Furthermore, when calculating the receiver’s position based on the selected mode, the proposed method achieves a root mean square (RMS) error of 16.87 m, representing a 4.7\% improvement over the 17.70 m RMS error of the existing method. 

\section{INTRODUCTION}

Urban environments present substantial challenges for positioning and navigation services reliant on the global navigation satellite system (GNSS) \citep{Lee22:Evaluation, Lee22:Urban, Zhang17:Current, Moon24:HELPS, Kim23:Low, Jia21:Ground}. 
Although GNSS is crucial to urban autonomous applications and intelligent transportation systems, dense infrastructure in these areas frequently obstructs, reflects, and diffracts GNSS signals \citep{Zhu18, Kim23:Machine, Kim23:Single, Lee23:Seamless}. 
These interactions reduce signal visibility and induce measurement errors, resulting in non-line-of-sight (NLOS) conditions and multipath effects that can degrade positioning accuracy by over 100 m in severe cases \citep{Wang12, MacGougan02}. 

To address these issues, 3D-mapping-aided (3DMA) techniques have been developed, utilizing high-accuracy 3D city models to improve GNSS reliability \citep{Kumar14, Adjrad17:Enhancing, Suzuki12:GNSS, vanDiggelen21:End}. 
A primary method within 3DMA is shadow matching, which compares observed GNSS signal behavior with predicted satellite visibility based on signal strength measurements \citep{Groves11:Shadow, Wang13:GNSS, Adjrad18:Intelligent}. 
Recently, zonotope shadow matching (ZSM), a set-based shadow matching technique, has emerged as a promising alternative to traditional grid-based shadow matching approaches \citep{Bhamidipati22:Set, Kim24:Performance}. 
The grid-based method discretizes the receiver position into predefined grid points, whereas the set-based method defines the receiver position and surrounding objects as geometric sets. 
In the set-based method, after the position estimation algorithm defines the final set area, the centroid of this area is utilized as the receiver position \citep{Bhamidipati22:Set}.
ZSM represents both the receiver position and buildings as set-based objects using constrained zonotopes. 
By leveraging the rapid vector concatenation capabilities of zonotopes \citep{Scott16:Constrained}, ZSM computes GNSS shadows—areas where line-of-sight (LOS) signals are obstructed by buildings—more efficiently, allowing for the calculation of a set-valued receiver position. 
This set-valued approach offers several advantages over grid-based methods, such as eliminating the trade-off inherent in discretization, where increasing accuracy by using finer grids also raises computational load. 
Furthermore, set-valued methods like ZSM facilitate direct safety checks, allowing for easier assessment of how well the estimated localization set aligns with user-defined safety boundaries, thereby improving reliability in urban positioning \citep{Bhamidipati22:Set}. 

Although ZSM efficiently estimates a set-valued receiver position using zonotopes, it still faces several challenges. 
One of these is the presence of multi-modal position ambiguity, which occurs not only in ZSM but also in conventional shadow matching \citep{Groves15:GNSS}. 
It represents a phenomenon where multiple receiver positions are estimated, leading to ambiguity in determining the correct receiver position. 
For instance, the set-valued receiver position estimated using ZSM arises as multiple disjoint sets, each referred to as a mode, rather than a single connected set. 
Selecting the correct mode is crucial for ensuring receiver position accuracy. 

To reduce multi-modal position ambiguity in ZSM, \cite{Neamati22:Set} proposed a method utilizing satellite-pseudorange consistency (SPC), which can be calculated using GNSS pseudorange measurements. 
This method involves constructing an SPC for each satellite to map the mode distribution in the 2D position domain to the range offset domain. 
Based on the SPC extracted from each satellite, a mixture model is generated, and the probabilities for each mode are calculated using this model. 
Then, the mode with the highest probability is selected. 
According to the previous study, a mode selection accuracy of 78\% was achieved when using the SPC with signals received in a single-timestep. 
The previous study also applied an iterative filtering approach to improve the performance of the SPC-based method by using multiple pseudorange measurements collected at a fixed location over an extended period. 
Additionally, a trained LOS classifier was developed to down-weight NLOS satellites, increasing the mode selection accuracy to 100\%. 
However, in practice, obtaining multiple pseudorange measurements at a single location in dynamic environments is challenging, and the performance of the trained LOS classifier can vary significantly depending on the surrounding environments. 
Therefore, algorithms are required to accurately select modes, even when only pseudorange measurements from a single-timestep are available and no trained LOS classifier is utilized. 

In this study, we propose a set-based position ambiguity reduction method for zonotope shadow matching in urban areas using estimated multipath errors. 
Unlike the existing method that creates SPC using pseudorange measurements contaminated by multipath errors, we aim to estimate and correct these errors to generate enhanced SPC. 
By leveraging the distribution of modes obtained from ZSM, we can predict the approximate propagation paths of satellite signals, allowing us to estimate the multipath errors. 

The remainder of this paper is organized as follows: Section \ref{sec:Existing} introduces the existing SPC-based mode ambiguity reduction method. Section \ref{sec:Method} presents our proposed method. Section \ref{sec:Field} evaluates the performance of our approach and compares it with an existing method. Finally, Section \ref{sec:Conclusion} concludes this paper. 


\section{Existing SPC-Based Mode Ambiguity Reduction Method} 
\label{sec:Existing}

This section introduces an existing mode ambiguity reduction method developed by \cite{Neamati22:Set}.
It consists of two steps: computing set-based SPC projections and identifying the most probable mode using an iterative set-based filter. 

\subsection{Computing Set-Based SPC Projections}
\label{sec:Computing}

\begin{figure}[H]
    \centering
    \includegraphics[width=0.8\linewidth]{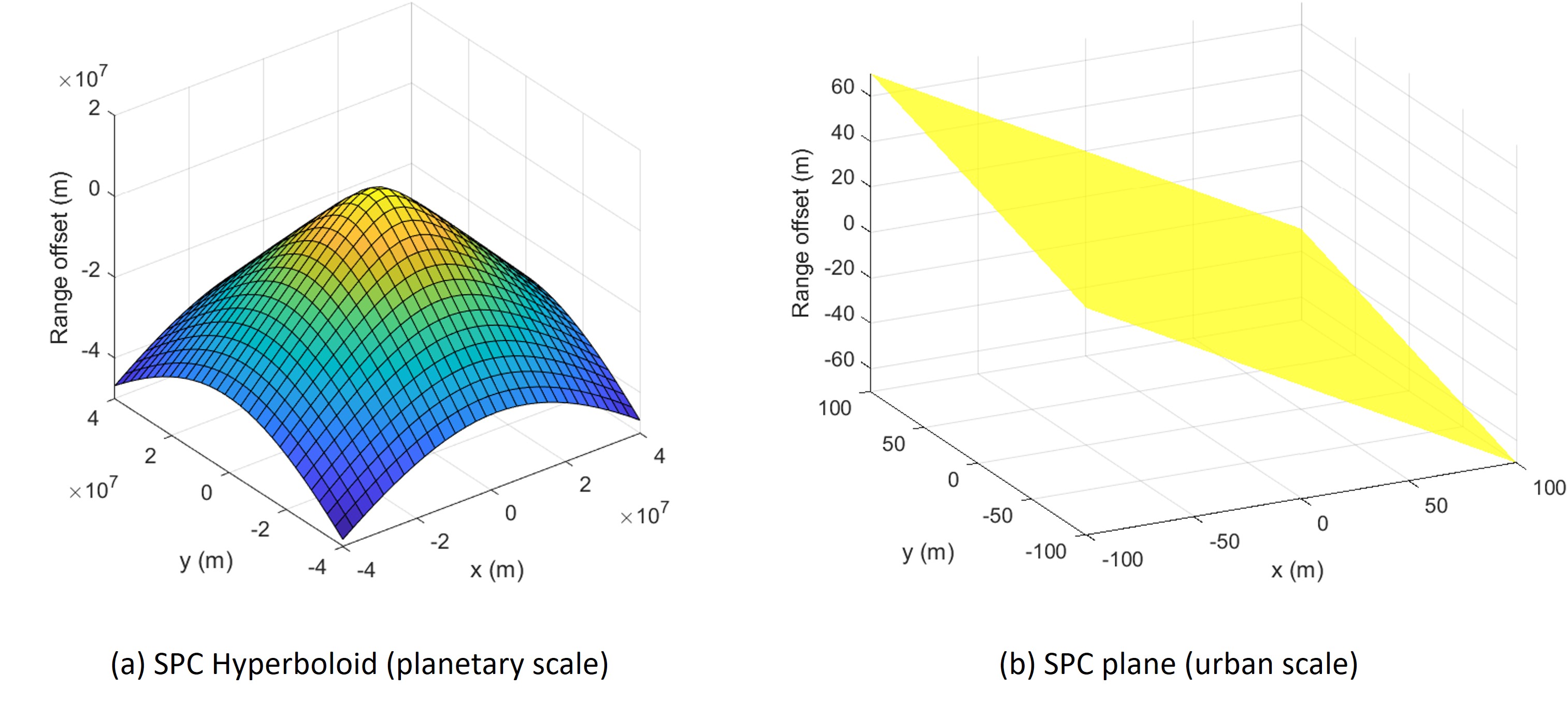}
    \caption{(a) The SPC hyperboloid at the planetary scale and (b) its plane-like appearance when observed at the urban scale (adapted from Figure 1 in \cite{Neamati22:Set}).}
    \label{fig:SPC}
\end{figure}

The SPC projection is defined as the transformation of the mode distribution, represented in the 2D position domain, into the range offset domain. 
This process can be applied to each satellite based on the following pseudorange formula: 
\begin{equation} \label{eq:Pseudorange}
\begin{split}
\rho^s &= \sqrt{(x^s-x^r)^2+(y^s-y^r)^2+(z^s-z^r)^2} + b_c + b_\mathrm{MP}^s + \epsilon
\end{split}
\end{equation}
where $\rho^s$ represents the pseudorange measurement corrected for atmospheric effects. 
$x^s$, $y^s$, $z^s$ and $x^r$, $y^r$, $z^r$ denote the position coordinates of the satellite and receiver, respectively. 
$b_c$ represents the receiver clock bias, $b_\mathrm{MP}^s$ indicates the multipath error, and $\epsilon$ represents thermal noise. 
Here, by substituting $b_c + b_\mathrm{MP}^s + \epsilon = b_\mathrm{ro}^s$, the equation is modified as follows:
\begin{equation} \label{eq:Pseudorange2}
\begin{split}
(x^s-x^r)^2+(y^s-y^r)^2+(z^s-z^r)^2-(\rho^s-b_\mathrm{ro}^s)^2 = 0
\end{split}
\end{equation}
In ZSM, it is assumed that the receiver is on the ground plane, so $(z^s-z^r)^2$ is fixed as a constant based on terrain information. 
Therefore, Equation (\ref{eq:Pseudorange2}) can be considered to consist of three unknowns: $x^r$, $y^r$, $b_\mathrm{ro}^s$. 
By representing this in the three-dimensional domain of ($x^r$, $y^r$, $b_\mathrm{ro}^s$), a hyperboloid, as shown in Figure~\ref{fig:SPC}(a), is obtained. 
When the range of the $x$ and $y$ axes is reduced to a small scale, as in an urban environment, the hyperboloid appears as a plane, as shown in Figure~\ref{fig:SPC}(b). 
To reduce computational complexity, this technique approximates the SPC hyperboloid as an SPC plane. 

The SPC plane allows for the transfer of information between the receiver position domain ($x^r$, $y^r$) and the range offset domain $b_\mathrm{ro}^s$. 
The detailed process of transforming between the receiver position domain and the range offset domain using the SPC plane is described in \cite{Neamati22:Set}.
The reason for transforming the position domain into the range offset domain is to analyze which mode exhibits the most consistent range offset for each satellite. 
The term $b_\mathrm{ro}^s$ consists of the receiver clock bias, multipath error, and thermal noise components; for satellite signals free from multipath contamination, this value will exhibit consistency. 
Therefore, by utilizing the SPC planes of each satellite, the consistency between the transformed range offsets can be evaluated, and the mode with the highest consistency can be identified as the most probable mode. 
The method for identifying this most probable mode will be explained in the following subsection. 

\subsection{Identifying The Most Probable Mode with an Iterative Set-Based Filter}
\label{sec:Identifying}
A framework for identifying the most probable mode is designed using a backward-forward architecture. 
Assuming there are $S$ satellites and $M$ modes, the process begins by assigning an equal initial probability of $1/M$ to each mode. 

In the backward pass, each of the $M$ modes for each satellite is projected onto the range offset domain using the SPC plane, resulting in following range offset intervals for each satellite: 
\begin{equation} \label{eq:Interval}
\mathcal{I}_s = \{\mathcal{I}_{s,m}\}^M_{m=1} = \{[\underline{a}_{s,m}, \overline{a}_{s,m}]\}^M_{m=1}
\end{equation}
Based on these interval collections, a multi-interval uniform distribution (MIUD) in the range offset domain can be generated, and the formula is as follows: 
\begin{equation} \label{eq:MIUD}
\begin{split}
u_s =
\begin{cases}
    0 & b_\mathrm{ro}^s \notin \mathcal{I}_s \\
    \frac{1}{\sum_{m=1}^M (\overline{a}_{s,m} - \underline{a}_{s,m})} & b_\mathrm{ro}^s \in \mathcal{I}_s
\end{cases}
\end{split}
\end{equation}
Using the MIUDs from all $S$ satellites, a mixture model is constructed to integrate information across all satellites. 
This consolidates the range offset domain information from each satellite into a single unified model, as illustrated in the following equation: 
\begin{equation} \label{eq:Mixture}
p\left( b_{ro} \mid \{s\}_{1:S}, \{m\}_{1:M} \right) = \eta_{\mathrm{LOS}} \sum_{s=1}^S p(s = \mathrm{LOS}) \, u_s
\end{equation}
where $p(s = \mathrm{LOS})$ represents the probability, obtained from the trained LOS classifier, that the signal from satellite $s$ is LOS. 
This aims to down-weight the impact of signals contaminated by multipath. 
Additionally, $\eta_{\mathrm{LOS}} = \left( \sum_{s=1}^{S} p(s = \mathrm{LOS}) \right)^{-1}$. 

This mixture model represents the probability distribution of the receiver’s range offset at the given epoch. 
Based on this information, the next forward pass step determines the posterior probability that each mode contains the receiver's location. 

During the forward pass, a random sample of $K$ points in range offset is generated from the distribution $p\left( b_{ro} \mid \{s\}_{1:S}, \{m\}_{1:M} \right)$. 
Using the Dirichlet distribution shown in the Equation (\ref{eq:Dirichlet}), the posterior probability that the receiver is in a given mode is calculated. 
\begin{equation} \label{eq:Dirichlet}
p\left( m \mid \{s\}_{1:S}, \{b\}_{1:K} \right) = \mathrm{Dir}(\alpha_1, \cdots, \alpha_M)
\end{equation}
where $\alpha_{1:M}$ represents the pseudocount, which is incremented by the weight $p(s = \mathrm{LOS})$ each time one of the $K$ range offset samples falls within the range offset interval for satellite $s$ and mode $m$, as shown in Equation (\ref{eq:Alpha}).
\begin{equation} \label{eq:Alpha}
\alpha_m \leftarrow \alpha_m + \eta_{\mathrm{LOS}} \sum_{s=1}^{S} \sum_{k=1}^{K} p(s = \mathrm{LOS}) \, I(b_k \in \mathcal{I}_{s,m})
\end{equation}
where the indicator function $I$ checks whether the sample $b_k$ is contained within the interval $\mathcal{I}_{s,m}$. 
Each $\alpha_m$ starts with an initial value of 1, and if there are pseudorange measurements from multiple timesteps at a fixed location, the Dirichlet distribution can be iteratively updated, converging over time. 

\section{Method} 
\label{sec:Method}

To achieve high mode selection accuracy, the traditional approach requires collecting pseudorange measurements over multiple timesteps at a fixed location and performing an iterative process, along with utilizing a trained LOS classifier to mitigate the effects of multipath-contaminated signals. 
However, as mentioned earlier, in dynamic environments, collecting multiple pseudorange measurements at a fixed location is challenging, and the performance of a trained LOS classifier can vary significantly depending on the surrounding environments. 

Therefore, we propose a method to estimate multipath errors for NLOS satellites and to correct pseudorange measurements using these estimates, thereby generating enhanced SPC and constructing a more reliable mixture model. 
This section will first provide an overview of the algorithm, then explain the method for estimating multipath errors using mode distribution, and finally describe the strategy for identifying the most probable mode. 

\begin{figure}
    \centering
    \includegraphics[width=1.0\linewidth]{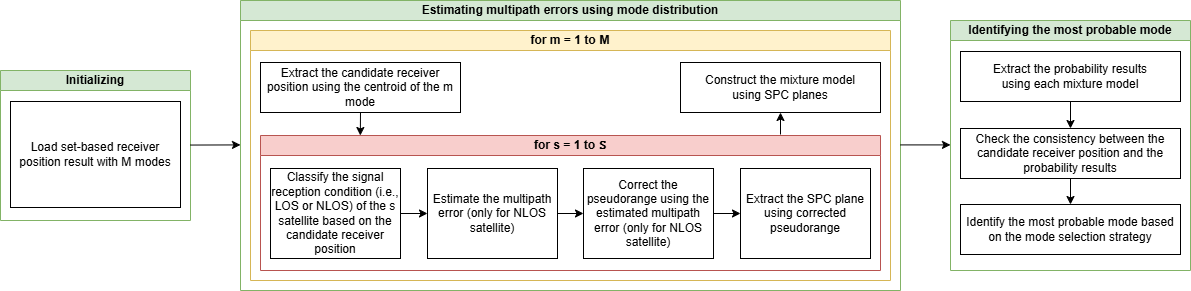}
    \caption{Flowchart of the proposed set-based position ambiguity reduction algorithm.}
    \label{fig:algorithm}
\end{figure}

\subsection{Algorithm Overview}
\label{sec:Algorithm}

Our set-based position ambiguity reduction algorithm consists of the process illustrated in Figure~\ref{fig:algorithm}. 
First, the set-based receiver position result is loaded. 
In this study, the set-based receiver position result was obtained as the output of the ZSM algorithm. 
Assuming there are $M$ modes and $S$ visible satellites, the multipath error estimation algorithm is applied to each mode, resulting in a total of $M$ mixture models. 
Details of the multipath error estimation algorithm are provided in Section~\ref{sec:Estimating}. 
Each mixture model result can be converted into a probability result, following the approach of the previous study explained in Section~\ref{sec:Identifying}. 
Note that, since an LOS classifier is assumed to be unavailable, the mixture model calculation in Equation~(\ref{eq:Mixture}) is modified to assign equal weights to all satellites, as shown below: 
\begin{equation} \label{eq:Mixture2}
p\left( b_{ro} \mid \{s\}_{1:S}, \{m\}_{1:M} \right) = \frac{\sum_{s=1}^S u_s}{S}
\end{equation}
The calculation formula for $\alpha_m$ in Equation~(\ref{eq:Alpha}) is also modified as shown below:
\begin{equation} \label{eq:Alpha2}
\alpha_m \leftarrow \alpha_m + \frac{\sum_{s=1}^{S} \sum_{k=1}^{K} I(b_k \in \mathcal{I}_{s,m})}{S}
\end{equation}
Additionally, assuming that pseudorange measurements are collected during a single-timestep at a fixed location, $\alpha_m$ is updated only once instead of iteratively over multiple timesteps. 
Using the $M$ probability results obtained, the most probable mode is identified, as explained in detail in Section~\ref{sec:Selecting}. 

\subsection{Estimating Multipath Errors Using Mode Distribution}
\label{sec:Estimating}

Since the mode distribution obtained from the ZSM contains receiver position information, the approximate signal propagation path from each satellite to the mode can be calculated. 
This information can then be used to estimate and correct multipath errors.

The multipath error estimation algorithm is performed for each mode, with the process as follows. 
First, the centroid of the mode is assumed to be the candidate receiver position. 
Based on this candidate position, the signal reception conditions (i.e., LOS or NLOS) for each satellite are classified. 
If a building obstructs the vector between the satellite and the candidate receiver position, the signal is classified as NLOS; otherwise, it is classified as LOS. 

\begin{figure}
    \centering
    \includegraphics[width=1.0\linewidth]{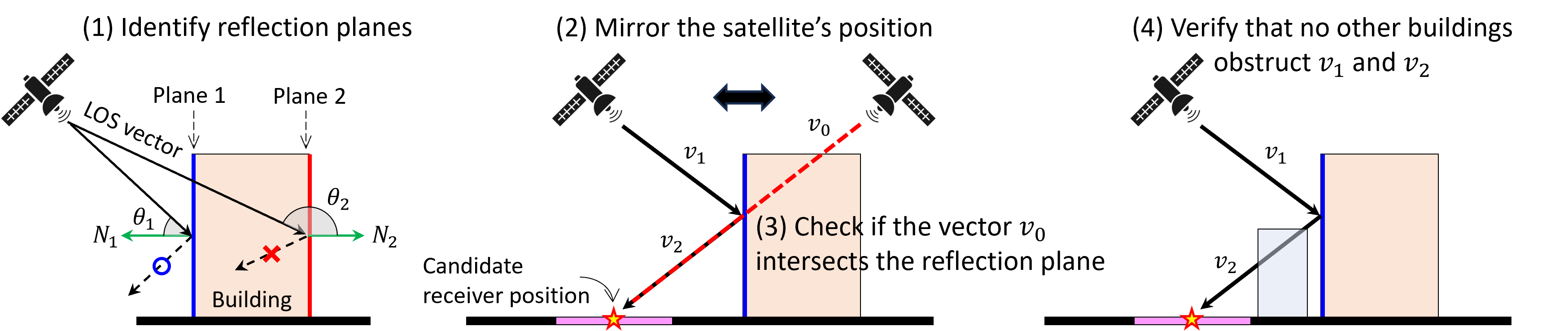}
    \caption{The process of identifying all possible propagation paths for a satellite.}
    \label{fig:CalcPropagationPath}
\end{figure}

Next, the propagation path delays are calculated for the NLOS satellites. 
It is assumed that the receiver receives only direct and single-reflected signals, ignoring those with multiple reflections. 
This assumption is reasonable, as receivers may not track multiple reflections due to their weak signal strengths. 
The process for identifying all possible propagation paths is as follows: 
\begin{enumerate}
  \item
  Identify reflection planes that could reflect the signal towards the ground plane. This can be determined based on the angle between the normal vector of each building plane and the LOS vector from the satellite toward the center of the plane, as shown in Figure~\ref{fig:CalcPropagationPath}. If the angle is less than 90 degrees, the plane is considered a reflection plane; otherwise, it is not. Additionally, cases where the reflected signal is directed toward the sky rather than the ground plane, such as the top surface of a building, are excluded from consideration as reflection planes. 
  \item
  Mirror the satellite’s position across the reflection plane. 
  \item
  Check whether the vector between the mirrored satellite and the candidate receiver position intersects the reflection plane. 
  \item
  If an intersection occurs, ensure that no other building planes obstruct the vectors connecting the receiver to the reflection plane or the satellite to the reflection plane. If unobstructed, store the propagation path for this reflection plane. Repeat Steps 2 through 4 for each reflection plane. 
\end{enumerate}

All the above processes were efficiently computed using the open-source Continuous Reachability Analyzer (CORA) toolbox \citep{Althoff15:An} by representing buildings and each vector as constrained zonotopes and performing set operations between the constrained zonotopes. 
Among all stored propagation paths, the shortest path is the first to reach the receiver. 
The multipath error is calculated based on this shortest path, defined as the difference between the computed propagation path distance and the measured pseudorange value. 
In some cases, a propagation path for a satellite signal may not be found. 
This could result from a discrepancy between the candidate receiver position and the true receiver position. 
If no propagation path exists, the multipath error calculation and correction processes are skipped. 

The pseudorange measurement is then corrected using the estimated multipath error, and an enhanced SPC plane is generated from the corrected pseudorange. 
For each satellite, an SPC plane is created, and a mixture model is derived from these SPC planes. 
This process is repeated for each mode, resulting in $M$ mixture models corresponding to the total number of modes. 

\subsection{Identifying The Most Probable Mode Strategy}
\label{sec:Selecting}

\begin{figure}[H]
    \centering
    \includegraphics[width=0.9\linewidth]{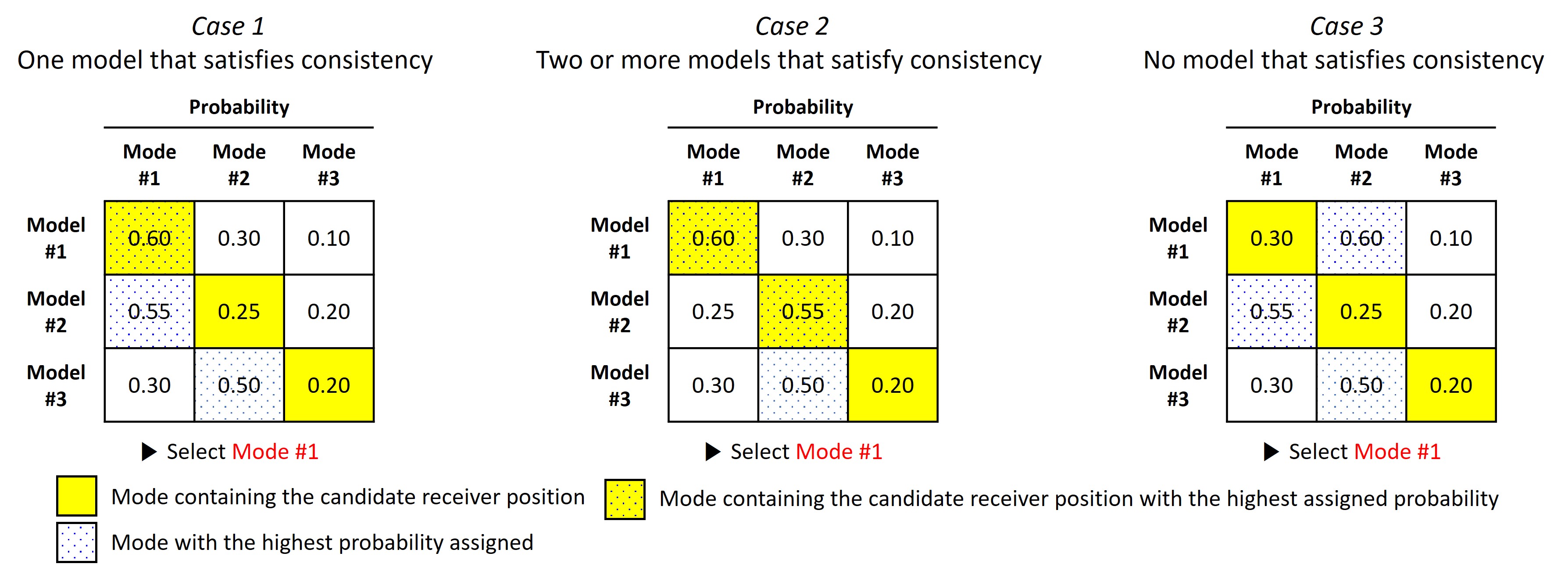}
    \caption{Proposed strategy for identifying the most probable mode by case type.}
    \label{fig:SelectModeStrategy}
\end{figure}

Using the constructed mixture models, the likelihood that each mode is the correct mode can be calculated. 
Since a total of $M$ mixture models were generated, $M$ probability results are obtained, which can be represented as an $M \times M$ array, as shown in Figure~\ref{fig:SelectModeStrategy} (example shown for $M$=3). 
To determine the most probable mode, we check the consistency between the mode with the highest probability in the mixture model and the mode corresponding to the assumed candidate receiver position used to generate the mixture model. 
For example, if the mixture model created using the multipath error estimated from the $m$\textsuperscript{th} mode assigns the highest probability to that mode, the result is considered reliable. 
Theoretically, there can be from 0 to $m$ mixture models that satisfy this consistency. 

As shown in Figure~\ref{fig:SelectModeStrategy}, we categorized the number of models satisfying consistency into three cases and defined a most probable mode selection strategy for each case. 
The first case is when only one model satisfies consistency. 
In this scenario, the most probable mode is selected based on the probability result of that model. 
The second case is when two or more models satisfy consistency. 
In this situation, the probability values of the modes assigned the highest probability by each mixture model are compared, and the mode with the highest probability is selected as the most probable mode. 
The last case is when no model satisfies consistency. 
In this case, a comparison is made between the probability values assigned to the modes that include the candidate receiver position for each model. 
For example, in case 3 of Figure~\ref{fig:SelectModeStrategy}, the comparison involves the values 0.30 for mode \#1 in model \#1, 0.25 for mode \#2 in model \#2, and 0.20 for mode \#3 in model \#3. 
Mode \#1, having the highest probability, is ultimately selected.

\section{Field Test Results}
\label{sec:Field}

For performance evaluation of the proposed method, we conducted a field test. 
The GPS data was collected in dynamic urban environments, and set-valued receiver position outputs were generated using the ZSM algorithm \citep{Bhamidipati22:Set}. 
We then compared the mode selection accuracy of our method to that of the existing SPC-based method. 
Additionally, we evaluated the receiver position accuracy based on the mode selection results. 

\subsection{Experimental Setup}

\begin{figure}[H]
    \centering
    \includegraphics[width=0.8\linewidth]{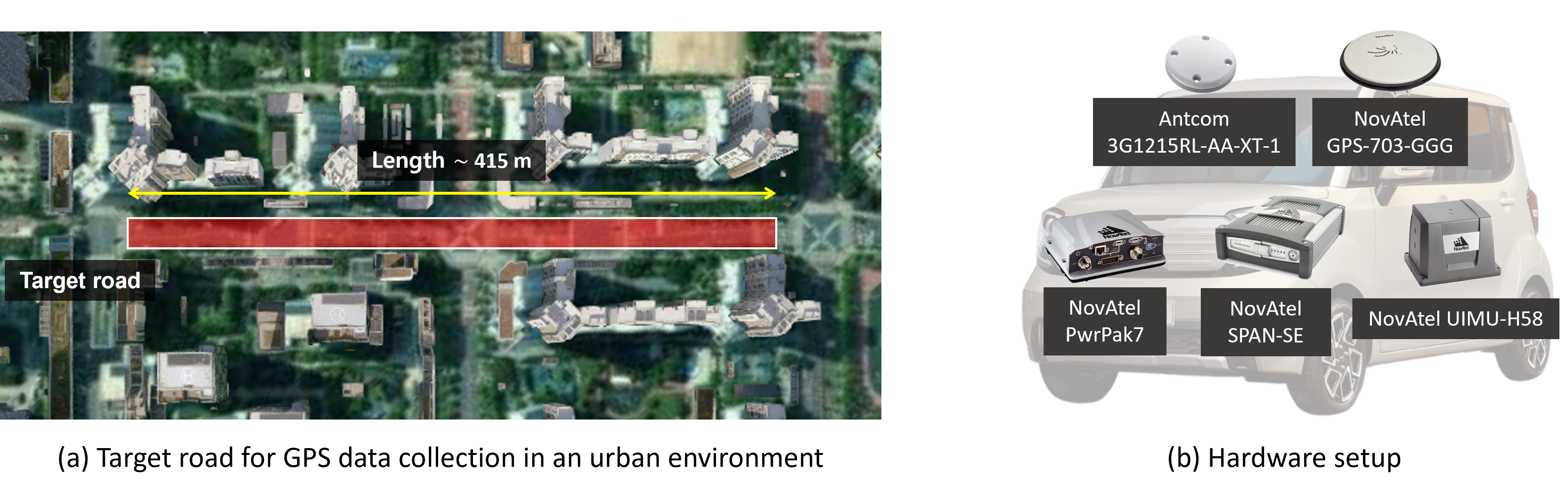}
    \caption{Target road and hardware setup for field test.}
    \label{fig:ExperimentalSetup}
\end{figure}

The GPS data collection experiment in an urban area was conducted on the target road shown in Figure~\ref{fig:ExperimentalSetup}(a), using the setup illustrated in Figure~\ref{fig:ExperimentalSetup}(b). 
We set up an Antcom 3G1215RL-AA-XT-1 antenna alongside a NovAtel PwrPak7 GNSS receiver on the vehicle, recording raw GPS data on a laptop at 1-second intervals throughout the experiment. 
To generate the true reference trajectory, the vehicle was also equipped with a NovAtel GPS-703-GGG antenna, NovAtel SPAN-SE, and NovAtel UIMU-H58. 
The GNSS/INS data were collected every second and later processed in Inertial Explorer using a tightly coupled mode to establish a precise ground truth trajectory. 

\subsection{Performance Evaluation}

\begin{figure}
    \centering
    \includegraphics[width=0.9\linewidth]{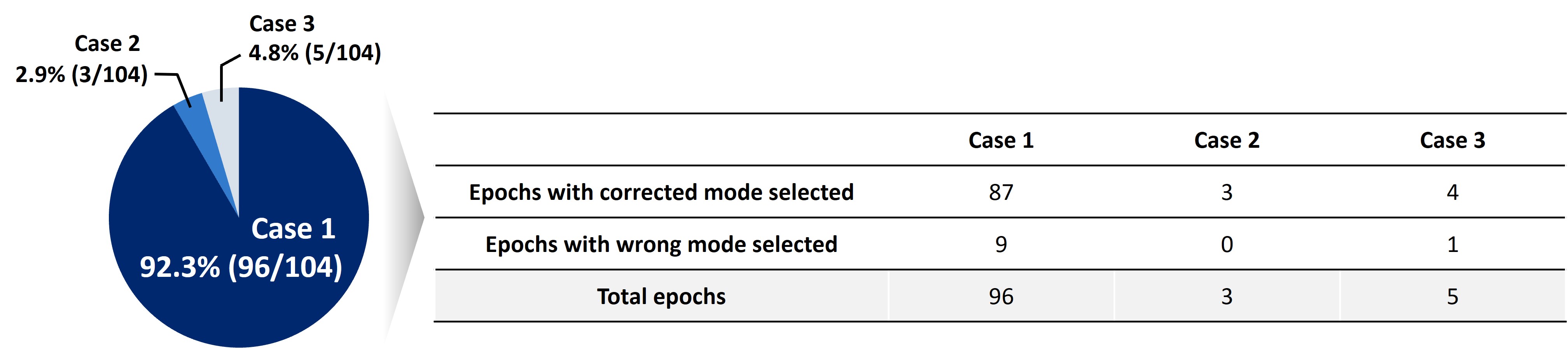}
    \caption{Occurrence rates of each case type in the proposed method, along with the number of epochs with correct and wrong mode selections for each case.}
    \label{fig:CaseOccurenceRate}
\end{figure}

To enable an equivalent comparison between the existing method and the proposed method, we assumed no trained LOS classifier was available and used only a single pseudorange measurement per satellite per epoch, rather than multiple measurements collected over several timesteps. 
As shown in Table~\ref{tab:sim_results}, the existing method selected the correct mode in 90 out of 104 epochs, achieving a mode selection accuracy of approximately 86\%.
In contrast, the proposed method achieved an accuracy of about 91\% by selecting the correct mode in 95 out of 104 epochs. 
The results of analyzing 104 epochs by the cases defined in Figure~\ref{fig:SelectModeStrategy} are shown in Figure~\ref{fig:CaseOccurenceRate}. 
Case 1 accounted for the majority with 96 epochs, followed by case 2 with 3 epochs and case 3 with 5 epochs. 
Regarding the mode selection results, 9 wrong mode selections occurred in case 1 and 1 in case 3. 

\begin{table}[H]
\small
\centering
\caption{Comparison of mode selection accuracy between the existing and proposed methods}\label{tab:sim_results}
\begin{center}{
\renewcommand{\arraystretch}{1.4}
 \begin{tabular}{ P{4cm} | P{3.5cm} | P{3.5cm}}
 \hhline{===}
 \rule{0pt}{15pt} \thead{Approach} & \thead{SPC \\ (Existing method)} & \thead{Enhanced SPC \\ (Proposed method)} \\
 \hline
 \rule{0pt}{20pt} \thead{Mode Selection \\ Accuracy} & 86\% (90/104) & 91\% (95/104) \\
 \hhline{===}
\end{tabular}}
\end{center}
\end{table}

\begin{figure}
    \centering
    \includegraphics[width=0.8\linewidth]{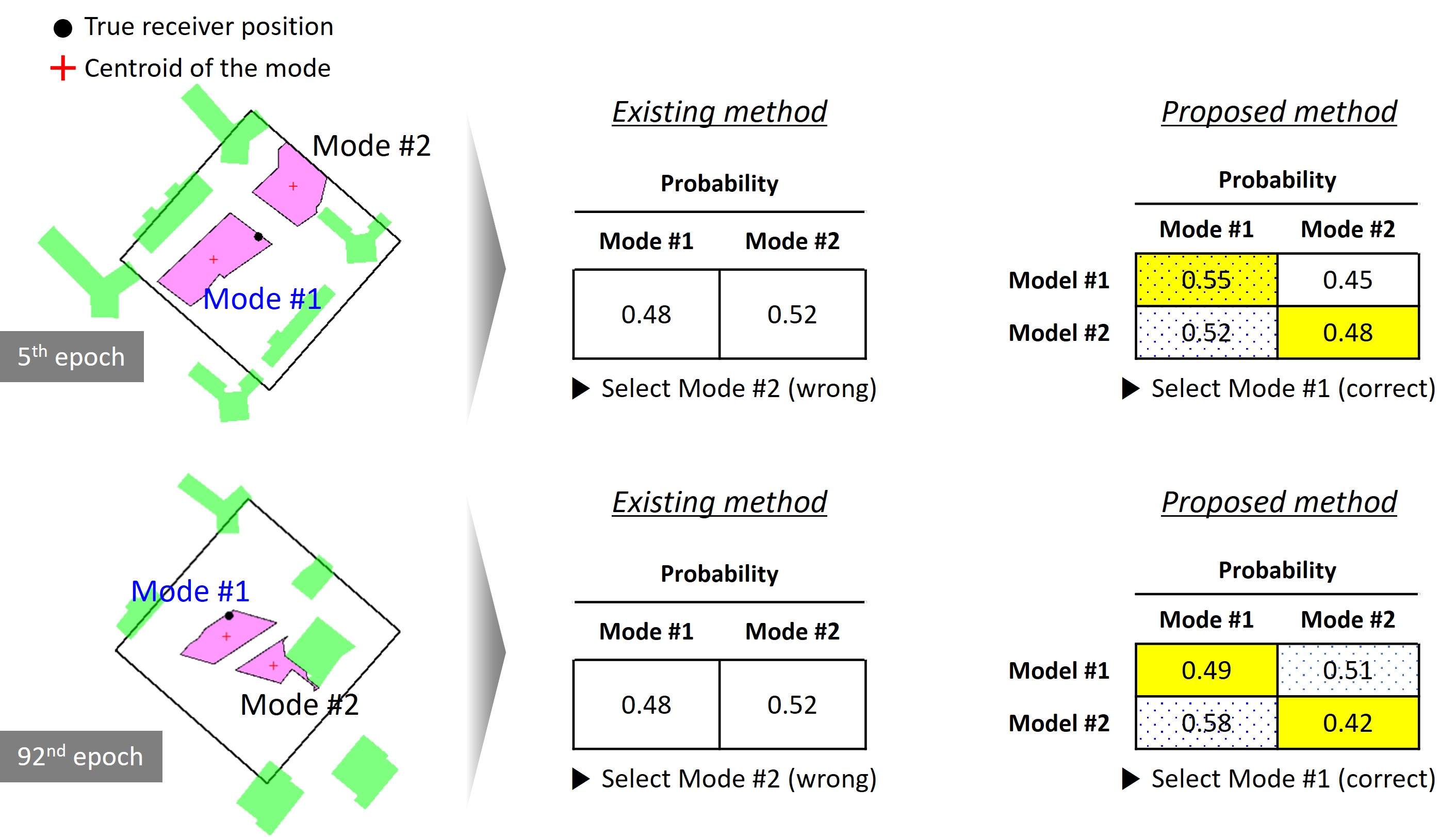}
    \caption{Mode selection results of the existing and proposed methods at the 5\textsuperscript{th} and 92\textsuperscript{nd} epochs.}
    \label{fig:Result1}
\end{figure}

Figure~\ref{fig:Result1} illustrates the mode selection results of existing and proposed methods at the 5\textsuperscript{th} and 92\textsuperscript{nd} epochs. 
In the case of the 5\textsuperscript{th} epoch, the ZSM algorithm produced two modes, with the true receiver position included in mode \#1. 
When the existing SPC-based method was applied, a probability of 0.52 was assigned to mode \#2, resulting in the selection of an wrong mode. 
In contrast, when the proposed method was applied, only model \#1 satisfied the consistency condition, leading to the selection of mode \#1 with an assigned probability of 0.55. 

In the 92\textsuperscript{nd} epoch, the existing method assigned the highest probability to an incorrect mode, resulting in the selection of a mode that did not include the receiver. 
When the proposed method was applied, neither model's probability results satisfied the consistency condition. 
Following the defined rule, the probability of mode \#1 from model \#1 was compared with that of mode \#2 from model \#2. 
As a result, the correct mode \#1, which included the receiver, was successfully selected. 

Furthermore, we analyzed the positioning accuracy of the ZSM algorithm based on the mode selection results. 
In this analysis, the receiver position was assumed to be the centroid of the selected mode. 
Ideally, if the mode containing the true receiver position were selected for all epochs, a root mean square (RMS) error of 14.02 m would be achieved. 
However, with the existing method, the RMS error increases to 17.70 m due to the occasional selection of a mode that does not include the true receiver position. 
In contrast, the improved mode selection accuracy of our method reduces the RMS error to 16.87 m, reflecting a 4.7\% improvement in positioning accuracy. 

\begin{table}[H]
\small
\centering
\caption{Comparison of positioning error between the existing and proposed methods}\label{tab:Error}
\begin{center}{
\renewcommand{\arraystretch}{1.4}
 \begin{tabular}{ P{3cm} | P{3.5cm} | P{3.5cm} | P{3.5cm}}
 \hhline{====}
 \rule{0pt}{15pt} \thead{Approach} & \thead{Ideal} & \thead{SPC \\ (Existing method)} & \thead{Enhanced SPC \\ (Proposed method)} \\
 \hline
 \rule{0pt}{20pt} \thead{RMS Positioning \\ Error} & 14.02 m & 17.70 m & 16.87 m \\
 \hhline{====}
\end{tabular}}
\end{center}
\end{table}

\section{Conclusion} \label{sec:Conclusion}

In this study, we propose a set-based position ambiguity reduction method for zonotope shadow matching, utilizing estimated multipath errors. 
By leveraging mode distribution information obtained from the output of ZSM, we predict the approximate signal propagation paths and estimate the multipath errors to correct the pseudorange measurements of NLOS signals. 
Furthermore, we derived an enhanced SPC plane using the corrected pseudorange and proposed a strategy for identifying the most probable mode. 
This approach enables us to achieve high mode selection accuracy without a trained LOS classifier, using only single-timestep pseudorange measurements. 
Evaluation using GPS data collected in an urban area demonstrated that the proposed method improved mode selection accuracy, achieving 91\% compared to 86\% for the existing approach. 
Furthermore, positioning accuracy was also enhanced with the proposed method, resulting in a 4.7\% reduction in RMS error to 16.87 m from 17.70 m. 

\section*{ACKNOWLEDGEMENTS}

This work was supported in part by the National Research Foundation of Korea (NRF), funded by the Korean government (Ministry of Science and ICT, MSIT), under Grant RS-2024-00358298; 
in part by the Future Space Navigation and Satellite Research Center through the NRF, funded by the MSIT, Republic of Korea, under Grant 2022M1A3C2074404; 
in part by Grant RS-2024-00407003 from the ``Development of Advanced Technology for Terrestrial Radionavigation System'' project, funded by the Ministry of Oceans and Fisheries, Republic of Korea;
in part by the Unmanned Vehicles Core Technology Research and Development Program through the NRF and the Unmanned Vehicle Advanced Research Center (UVARC) funded by the MSIT, Republic of Korea, under Grant 2020M3C1C1A01086407;
and in part by the MSIT, Korea, under the Information Technology Research Center (ITRC) support program supervised by the Institute of Information \& Communications Technology Planning \& Evaluation (IITP) under Grant IITP-2024-RS-2024-00437494.

\bibliographystyle{apalike}
\bibliography{mybibfile, IUS_publications}

\end{document}